\documentclass{vgtc}                          

\ifpdf
  \pdfoutput=1\relax                   
  \usepackage{graphicx}                
  \DeclareGraphicsExtensions{.pdf,.png,.jpg,.jpeg} 
\else
  \usepackage{graphicx}                
  \DeclareGraphicsExtensions{.eps}     
\fi%

\graphicspath{{figures/}{pictures/}{images/}{./}} 

\usepackage{microtype}                 
\PassOptionsToPackage{warn}{textcomp}  
\usepackage{textcomp}                  
\usepackage{mathptmx}                  
\usepackage{times}                     
\usepackage{cite}                      
\usepackage{tabu}                      
\usepackage{booktabs}                  

\onlineid{0}

\vgtccategory{Research}

\vgtcinsertpkg

\title{Understanding User Experience of COVID-19 Maps through Remote Elicitation Interviews}

\author{Damla \c{C}ay\thanks{e-mail: dcay13@ku.edu.tr}\\ %
        \scriptsize Ko\c{c} University %
\and Till Nagel\thanks{e-mail: t.nagel@hs-mannheim.de}\\ %
     \scriptsize Mannheim University \\ 
     \scriptsize of Applied Sciences %
\and As{\i}m Evren Yanta\c{c}\thanks{e-mail: eyantac@ku.edu.tr}\\ %
    \scriptsize Ko\c{c} University
     \parbox{1.4in}{\scriptsize \centering}}

\abstract{During the coronavirus pandemic, visualizations gained a new level of popularity and meaning for a wider audience. People were bombarded with a wide set of public health visualizations ranging from simple graphs to complex interactive dashboards. In a pandemic setting, where large amounts of the world population are socially distancing themselves, it becomes an urgent need to refine existing user experience evaluation methods for remote settings to understand how people make sense out of COVID-19 related visualizations. When evaluating visualizations aimed towards the general public with vastly different socio-demographic backgrounds and varying levels of technical savviness and data literacy, it is important to understand user feedback beyond aspects such as speed, task accuracy, or usability problems. As a part of this wider evaluation perspective, micro-phenomenology has been used to evaluate static and narrative visualizations to reveal the lived experience in a detailed way. Building upon these studies, we conducted a user study to understand how to employ Elicitation (aka Micro-phenomenological) interviews in remote settings. In a case study, we investigated what experiences the participants had with map-based interactive visualizations. Our findings reveal positive and negative aspects of conducting Elicitation interviews remotely. Our results can inform the process of planning and executing remote Elicitation interviews to evaluate interactive visualizations. In addition, we share recommendations regarding visualization techniques and interaction design about public health data.
} 

\CCScatlist{
  \CCScatTwelve{Human-centered computing}{Visu\-al\-iza\-tion}{Visualization design and evaluation methods}{}
}

\begin{document}



\maketitle

\section{Introduction}
Data visualization is becoming an important part of daily life as more and more people commonly encounter data-driven online platforms and tools in contexts ranging from journalism to urban spaces \cite{bird2010pulse,nagel2020visually,pousman2007casual,ccay2019happening}. The SARS-CoV-2 pandemic further increased this importance by putting data visualization in the center of everyday life~\cite{boulos2020geographical,worldhealthorganization,johnshopkins}. In this urgent situation, many organizations started to create visualizations to better explore and communicate the global effects of the pandemic. This created many challenges including designing visualizations with complex data for wide audiences, including experts, policymakers, and lay people with different levels of data literacy.

To understand the impact of these visualizations, it becomes critical to investigate the experiences people have using them while the pandemic is still ongoing. People use these visualizations to learn more about the subject, and inform their decisions about their personal life. Additionally, it is a topic that has a strong emotional aspect. These conditions put COVID-19 visualizations in a unique position where there is an opportunity to learn interesting and relevant information about how people interact with these visualizations in the course of their everyday life, and learn from their experience to drive visualization design decisions accordingly. 

For a phenomenon that affects many people deeply on a personal level, we believe it is important to investigate user experiences beyond performance measures to capture the richness of their experience. Traditionally, evaluation methods for visualization are mostly based on usability and performance ~\cite{lam2011empirical,isenberg2013systematic}. Recent studies have been introducing methods from other fields to capture the richness of the user experience further~\cite{brehmer2013multi,haroz2015isotype,kennedy2016engaging,hogan2015elicitation}. Hogan et al.~\cite{hogan2015elicitation} introduced Micro-Phenomenological Interviews for visualization studies and described it as the Elicitation Interview (EI), that aims to capture personal experiences with great detail and with less risk of bias and post-rationalization~\cite{petitmengin2013gap}. They suggest that visualizations can trigger associations with past experiences, therefore affect the user’s experience heavily. We believe COVID-19 visualizations are especially good cases for EI as the reading and usage might be strongly influenced by the user's personal experience and knowledge. 

To understand the user experience with data visualizations during the pandemic, we need to adapt co-located evaluation methods for remote settings, due to the existing mobility disruptions, and because a large portion of the population is social-distancing or even quarantined at home. In the last decade, researchers started to conduct remote interviews to understand their advantages and disadvantages over physical interviews. Remote interviews have many advantages in terms of reducing cost and increasing accessibility by removing geographical and mobility constraints~\cite{janghorban2014skype}. On the other hand, challenges were reported regarding connectivity and rapport ~\cite{seitz2016pixilated}. Poor internet connection can cause calls or recordings to be disrupted ~\cite{brown2018video}. While studies found remote interviews supportive for building rapport, they caution against a deceptive closeness between the participant and researcher ~\cite{brown2018video}, and point out that the lack of intimacy in remote interviews may cause problems for sensitive questions ~\cite{seitz2016pixilated}. More recent literature reports that remote interviews resemble in-person interviews in rapport~\cite{jenner2019intimacy}. Remote interviews in the literature span all types of interviews from structured, to semi-structured, to open-ended interviews. We are not aware of any studies that applied the Elicitation interview technique remotely.

In this paper, we describe how we conducted a remote user study on two COVID-19 visualizations with seven participants. We selected the WHO Coronavirus Disease (COVID-19) Dashboard~\cite{worldhealthorganization} and the COVID-19 Dashboard by the Center for Systems Science and Engineering (CSSE) at Johns Hopkins University (JHU)~\cite{johnshopkins}, as both are from reputable sources, present similar types of information, have comparable views, yet use distinct visual styles. In the literature of elicitation interviews used for data visualizations, the method has been applied in a co-located setting with static visualizations\cite{hogan2015elicitation}. Nowak et al.~\cite{nowak2018micro} used the EI method for narrative visualizations and conducted two of the six interviews remotely, yet did not discuss the challenges or advantages. We aim to build upon these previous studies on Elicitation Interviews by expanding the method to a remote setting and applying it to interactive visualizations. With our study, we contribute considerations on how to conduct remote elicitation interviews for interactive visualizations and design recommendations for public health visualizations.


\section{Background}
As visualizations become more important for our daily lives, evaluating and understanding how people interact with these visualizations becomes more critical to create better visualization tools. Previous studies have extensively analyzed different evaluation methods for different types of visualization scenarios ~\cite{lam2011empirical, isenberg2013systematic}. The visualization literature has a growing pool of evaluation methods for different visualization stages and purposes ~\cite{munzner2009nested, lam2011empirical, isenberg2013systematic}. Yet, visualization evaluation studies still mostly focus on user performance, through assessing objectively measured metrics like time on task and accuracy ~\cite{hogan2015elicitation, lam2011empirical, isenberg2013systematic}. 

Some more recent studies move beyond performance-related objectives, including metrics like enjoyment ~\cite{brehmer2013multi}, engagement ~\cite{haroz2015isotype, kennedy2016engaging}, memorability ~\cite{saket2015map, borkin2015beyond}, and emotion~\cite{harrison2013role, wang2019emotional}. Wang et al.~\cite{wang2019emotional} argue that researchers typically concentrate on performance, comprehension, or insights when determining the importance of visualizations. In their work, they expand these definitions by emphasizing emotional responses and including different forms of engagement. They propose an alternative way to value data representations: by their creativity and their ability to engage. They expand engagement further by defining different dimensions of it, which are affective engagement, physical engagement through touch and movement, intellectual engagement, and social engagement. In another study focused on visualization engagement, Kennedy et al.~\cite{kennedy2016engaging} explored subjective indicators of engagement through a diary study and a set of focus groups. Some of the indicators were provoking questions, creating empathy, generating curiosity, provoking surprise, changing minds, or provoking a strong emotional response.

There are different methods and metrics to get a deeper understanding of the user’s experience of interacting with a visualization. One of the most common methods is the interview method, which is generally structured, semi-structured or open-ended interviews with a focus on the usability aspects ~\cite{lam2011empirical}. These types of interviews fall short when we want to get a deep understanding of the user’s experience at the moment of using the visualization, as they carry the risk of post-rationalization ~\cite{hogan2015elicitation}. To overcome this, recent studies used the Elicitation Interview technique on visualizations ~\cite{hogan2015elicitation, nowak2018micro}. It is a method of descriptive phenomenology, developed by the psychologist Vermersh ~\cite{vermersch2019entretien}, as he was aiming to transfer the implicit knowledge of experts to non-experts. Later on, this method was used with any type of lived experience by the cognitive scientist Varela ~\cite{petitmengin2006describing}. The method was also used in the HCI field for understanding text entry experiences ~\cite{light2008transports}, tactile experiences ~\cite{obrist2013talking} and recently in the visualization field to understand the experience of making sense of data displays ~\cite{hogan2015elicitation, hogan2017visual, nowak2018micro}. For static and narrative visualizations, the method was found useful as it allows understanding users’ subjective experiences beyond their attitudes and judgments ~\cite{hogan2015elicitation, nowak2018micro}. In these studies, Hogan et al.’s ~\cite{hogan2015elicitation} focus was on static visualizations while Nowak et al.’s ~\cite{nowak2018micro} focus was on narrative visualizations with limited interactivity. However, there have been no studies on more complex interactive visualizations. We believe applying EI to complex interactive visualizations is important, as interactivity is an important element for visualizations.

The importance of remote methods increased as the COVID-19 Pandemic limited the mobility of societies worldwide. Before this situation, research on remote interviews gained speed with the rapid developments in technology and the wide usage of available tools. Online interviews have an immense strength that overcomes temporal, financial, geographical, and mobility constraints~\cite{janghorban2014skype}. As a result, remote interviews contribute to inclusion and diversity of participants~\cite{iacono2016skype}. However, they introduce new types of challenges. One important challenge is connectivity problems~\cite{brown2018video}. Even though connectivity problems decreased significantly over the years of remote interview literature, even small problems can still affect the flow state negatively for EI. Another challenge identified by several studies~\cite{seitz2016pixilated, brown2018video}, is related to rapport. Seitz~\cite{seitz2016pixilated} notes that it is hard to build rapport in remote interviews as there is no physical,  in-person relationship. On the other hand, Brown~\cite{brown2018video} states that remote interviews can create a deceptive closeness, as the tools used (e.g. Skype) are usually used with family and friends. We believe this perception might be shifted due to the widely applied remote working situations. Additionally, a more recent study by Jenner et al.~\cite{jenner2019intimacy} reports that remote interviews resemble in-person interviews in rapport and the depth of information shared. 

Hogan et al.~\cite{hogan2015elicitation} uses the Elicitation Interview in a co-located setting and even described how the physical adjustments should be to assist the desired mental state. Nowak et al.~\cite{nowak2018micro} conducted two of the six interviews in a remote setting, however, they note that it was done for convenience and did not discuss it further. We believe it is important to understand and highlight the advantages and challenges of remote elicitation interview methods on interactive visualizations.

\section{METHODOLOGY: Elicitation Interviews}

\begin{figure*}
\includegraphics[width=\textwidth]{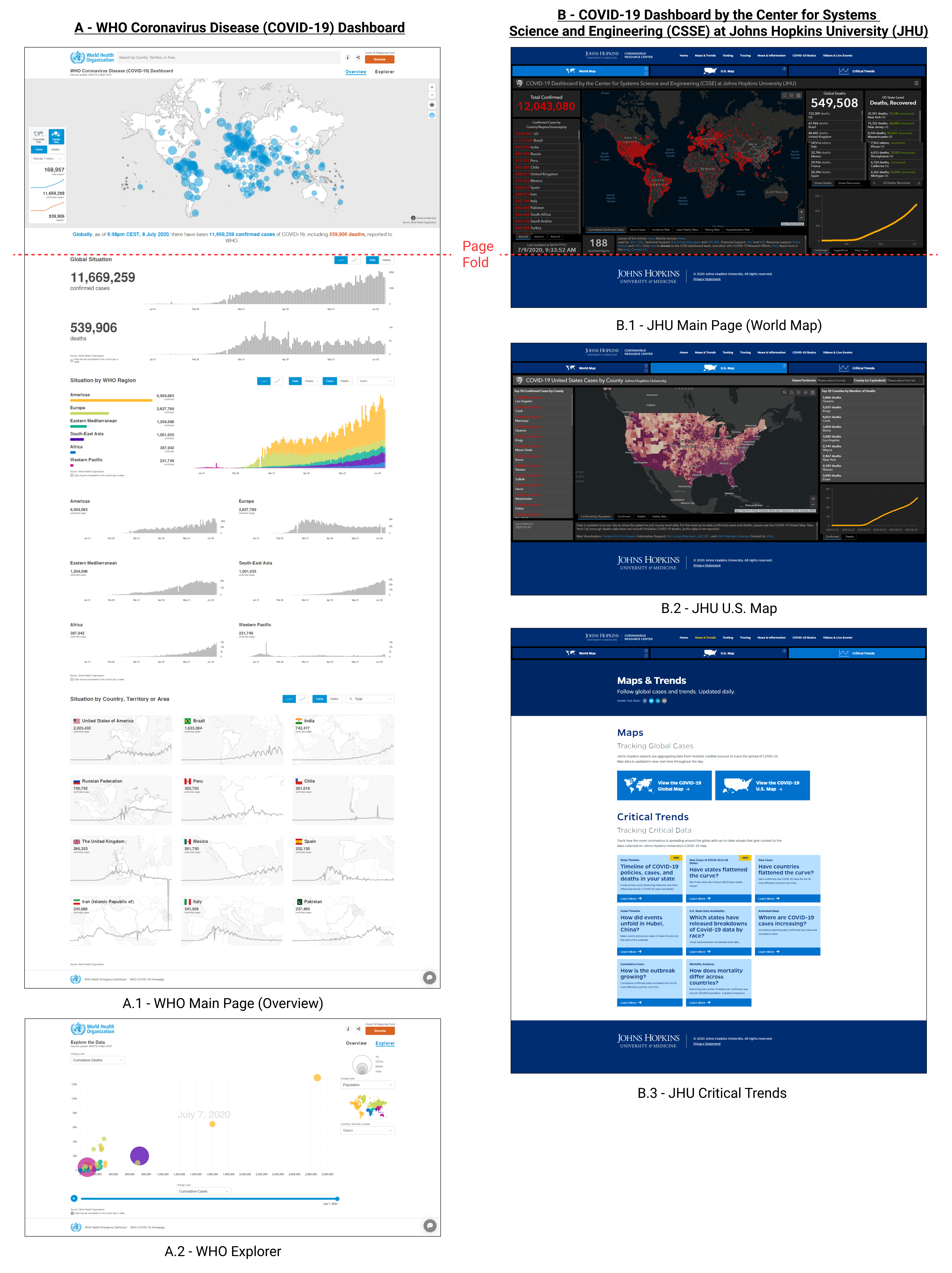}
  \caption{Main pages of two websites that were explored by the participants. A.WHO Coronavirus Disease (COVID-19) Dashboard~\cite{worldhealthorganization} B.COVID-19 Dashboard by the Center for Systems Science and Engineering (CSSE) at Johns Hopkins University (JHU)~\cite{johnshopkins} Red dashed line indicates the page fold of the main pages of both websites.}
  \label{fig:Frame5}
\end{figure*}

An elicitation interview requires a mental state that enables the evocation of the lived experience. The participant should feel as if they were living that moment. This state enables acquiring details that would otherwise go unnoticed. To create such a mental state, an elicitation interview uses a special questioning technique and interview structure. The interview is conducted in a quiet place without any distractions. During the interview, open-ended and content-empty questions are asked in the present tense, to evoke the experience. The questions are “What” and “How” questions, no “Why” questions are asked to avoid judgments and post-rationalization. The interviewer observes specific cues to understand if the participant is in an evocation state. Speaking in the present tense, gazing into space, taking pauses, and slurring speech are some of the cues that indicate the participant is in the evocation state~\cite{petitmengin2006describing}.

We primarily use Hogan et al.’s~\cite{hogan2015elicitation} detailed elicitation interview guide where the phases and the question types are explicitly framed for visualization studies. The questioning technique has an iterative structure, the level of detail increases through the interview. The questions themselves are content-empty, they don’t introduce new information, but they aim to reveal the actions, sensations, and feelings of the participant at an increasing level of granularity. An important pre-interview consideration is selecting a particular experience. The elicitation interview requires focusing on a singular experience, it is not suitable for general experiences. The common case for visualization studies has been creating that experience by providing users with a visualization to view, either a week before the interview~\cite{hogan2015elicitation}, or right before the interview~\cite{hogan2017visual, nowak2018micro}. 

The interview structure has six phases. The first phase is notifying the participant about the aim of the study, and the nature of the interview, that it can go deeper into experiences than regular interviews. The participant should be aware that they can stop the interview at any point. The second phase is commencing the interview  using the phrase, “If you agree, I would like to go back to the time when you started to experience X.” Then the questions of when and where the experience happened are asked. After this, the third phase begins where the evocation state is induced through visual, auditory, kinesthetic, and emotional questions. Some examples of these questions are: “When you are there, what do you see? At this moment, what are you hearing? At that time, what is the position of your body?, At that time, what are you feeling?”. The fourth phase is called the diachronic stage. It aims to understand the temporal structure of the experience, how it unfolded over time.  Example probing questions that mark the key moments of the experience are: “How do you start?, What do you do then?,  What happens at the end?, How do you know you have finished?”~\cite{petitmengin2009validity}.

After the temporal structure is defined, the fifth phase begins. It is called the synchronic structure and it aims to deepen the episodes defined in the previous phase. Thus, probing questions are more fine-tuned versions of the evocation stage (third stage) questions, that aim to reveal visual, auditory, kinesthetic, and emotional sensations in greater detail. At the sixth and last phase, the interviewer reformulates the entire interview and asks the participant if they have anything to add. 

Our work expands the previous works on elicitation interviews by applying the method in a remote context. In addition to the elicitation interview about the experience, we asked post-interview questions to understand the implications of the method being remote. We discuss how the non-verbal cues that indicate an evoked state is translated to a remote context. Furthermore, the visualizations we selected are tools where the users can freely explore the collection of visual information. This creates a challenge of understanding the navigation behavior precisely since there are no visual records but only the remarks of the user. We believe that elicitation interviews can still be useful without a visual record, and we discuss the level of insights that this type of information can induce. We see our work as a probe investigating the elicitation interview methodology in a remote setting, and intend this paper to function as a case study. Through describing our methodologies and reflecting on what worked and what did not, we hope to provide some insightful lessons for others to plan and conduct their elicitation interviews remotely, and to benefit future studies.

\subsection{Participants and Procedure}

Seven participants (5 female, 1 male, 1 prefers not to say) participated in the study. The range of their ages was between 22 and 43 (Median: 25). Participants rated their experience with data visualizations and introspective practices from 1 to 5 (1 Not experienced - 5 Very experienced). On average, participants rated 2.8 (SD: 1.6) for their data visualization experience and 3 (SD: 1.5) for their introspective experiences. The participants had diverse professions, including a florist, an industrial engineer, a mechanical engineer, a clinical psychologist, an architecture student, a human-computer interaction researcher, and an academic. The interviews were conducted in June 2020. Five of the participants were located in Istanbul, Turkey and two of them were located in Glasgow, UK. Pandemic measures for both countries were started to ease down for both countries starting from June.

For recruitment, we created an open call through our research group’s Instagram page. We sent a survey to the participants and asked them to complete the survey from their laptop or desktop computers before the video call. The survey had questions about their background, experience about their introspective practices (meditation, yoga, mindfulness, and cognitive behavioral therapy), and data visualization. We asked their experience about introspective practices because, in previous micro-phenomenological studies, researchers found that participants with experience in introspective practices were more successful in focusing on, and communicating their inner thoughts, actions, and feelings ~\cite{nowak2018micro}. After these questions, the survey presented the link of the first visualization website and asked the participants to explore it until they felt they gained as much information as possible. As the last step of the survey, the participants explored the second website. We randomized the shown order of websites for each participant to balance order bias.

We selected two visualizations to be used by participants. The first one is the World Health Organization’s (WHO) COVID-19 map visualization ~\cite{worldhealthorganization}. The second one is Johns Hopkins University’s COVID-19 map visualization ~\cite{johnshopkins}. We focused on these visualizations as the topic is urgent and the previous studies haven’t explored this method with exploratory, interactive visualizations. We selected these specific visualizations as they were both from reputable and well-known sources, both included similar types of information, both had a prominent spatial visualization, yet two websites differed in terms of visual style. WHO Coronavirus Disease (COVID - 19) Dashboard has a choropleth map with blue shades on a light grey background (See Figure ~\ref{fig:Frame5}.A). On the bottom, a label displays total confirmed cases and total deaths. On the left, a small widget allows the user to switch to a graduated circle map (referred as bubble map in the website) view, or switch to see deaths instead of cases. Small line charts show new cases and confirmed cases over time. The map is not scrollable with a mouse. When the user scrolls down, there are bar graphs of confirmed cases over time, for continents and different countries. Johns Hopkins University’s COVID-19 Map has a bubble map with red circles on a dark grey background (See Figure ~\ref{fig:Frame5}.B). On the left side, there is the number of total confirmed cases and a list of confirmed cases by countries. On the right side, there is the number of global deaths and a list of deaths per country. Next to it, deaths and recovered are listed at the US State level. On the right bottom, there is a line chart of total confirmed cases by time. 

We used Zoom for the video call, a free video-conferencing tool with a recording option~\cite{zoom}. When planning the video call, we asked participants to select a time and space where they can be away from distractions. The interview started with a brief introduction to the study. We followed Hogan et al.’s interview guide~\cite{hogan2015elicitation}. After the elicitation interview finished, we asked post-interview questions about their general experience during the pandemic process, their experience with remote activities, and what they think about the interview being remote.

\subsection{Data Analysis}
For analyzing the EI data, we used thematic analysis as suggested by previous studies ~\cite{hogan2015elicitation, nowak2018micro}. Our approach included three stages, (1) familiarization, (2) thematic coding, and (3) creating and refining themes. At the first stage, the researcher who conducted the interview also transcribed them to familiarize themselves with the data. The same researcher coded the data using open coding. After the primary researcher finished the initial coding, another researcher reviewed them and the codes were revised after discussions. The same two researchers collaboratively created themes from the codes. We used the remote collaboration tool Miro~\cite{miro} for this step. After three iterations, we defined and named the four final themes.

\section{Results}

\begin{table*}[ht]
\caption{List of themes emerged from the thematic analysis of the remote elicitation interview. Numbers in parenthesis indicate the number of participants the code was applied.}
\centering
\begin{tabular}{p{0.25\linewidth}p{0.25\linewidth}p{0.25\linewidth}}
\hline
Themes & Description & Illustrative Quote\\
\hline
\\\\
\textbf{Visual Impressions:} \\
Color (7), Layout (5), Ambiguous details (5)
 & Experiences and feelings that were influenced by the visual variables like color, general layout or interface elements 
 & \textit{“It's all red on black, the colors make me feel like the world is getting worse.”}[P4, JHU](Fig ~\ref{fig:Frame5}.B.1) \\\\
 \hline
\\\\
\textbf{Interactive Exploration:}\\
Spatial Exploration(4), Engaging with text (3), Previous knowledge (4), Completing Exploration (5)
 & Experiences of open-ended exploration processes 
 & \textit{“I have a planned flight for a wedding so I’m looking at both countries’ numbers.”}[P1, WHO](Fig ~\ref{fig:Frame5}.A) \\\\
 \hline
\\\\
\textbf{Information Overload:} \\
Refraining Depth (7), Losing meaning (5), Feelings about subject (3)
 & Feelings caused by the amount, visual form, or content of data 
 & \textit{“I feel like I can go on forever on this website. I don't want to confuse myself so stay on the main page and don't search anything.”}[P3, WHO](Fig. ~\ref{fig:Frame5}.A.1) \\\\
 \hline
\\\\
\textbf{Expectations of Functionality:} \\
Map (5), General Functionality (4)
 & Experiences of functionality expectations that are not being met and causing negative feelings like boredom and frustration 
 & \textit{“I wanted to see more layers on map but it wasn't working the way I wanted, then I experimented with the U.S. Map page(Fig ~\ref{fig:Frame5}.B.2)but I was bored easily because of my disappointment with the main page(Fig ~\ref{fig:Frame5}.B.1)”}[P5, JHU]\\\\

\hline
\end{tabular}
\end{table*}

In this section, first, we will present the themes that emerged from the remote EI. Then we will present our observations from the interviews and results of the post-interview questions about participants’ remote social experiences and their relationship with data during the COVID-19 pandemic. We will use the abbreviations “WHO” for World Health Organization’s Coronavirus Map, and “JHU” for Johns Hopkins University’s website while presenting our results. We will use identifiers in square brackets for distinguishing participants' statements. 

\subsection{Elicitation Interview Themes}

We present findings of how users explore a map, interact with rich data, and complete their exploration. The interviews also revealed issues that could be improved in terms of user experience. We share our results building on the findings of Hogan et al. and Nowak et al. Some of the themes that emerged from applying EI to static and narrative visualizations were also evident in our study. More specifically, mentioning colors, comparing data with previous knowledge, finishing viewing when one has a general understanding, and finding personal connections were common points with previous results on static and narrative visualizations. 

\subsubsection{Visual Impressions}
This theme includes participants’ common impressions regarding visual elements such as the use of color, general layout, and interface elements. In all interviews, color was the first visual variable to be mentioned for both websites. For WHO, the color blue and white were noticed first. The color theme was perceived positively by all participants. Participants described the colors as nice and pleasant. One participant noted that WHO does not have “a crisis feeling”[P2]. On the other hand, for JHU, participants noted that it is pessimistic, it shows danger, and it’s tiresome. One participant noted, \textit{“It's all red on black, the colors make me feel like the world is getting worse.”}[P6]. This theme is similar to the “Interpretation Processes of Visuals” theme from Hogan et al.’s ~\cite{hogan2015elicitation} work and “Experiences of Aesthetics and Design” theme from Nowak et al.'s work ~\cite{nowak2018micro}. Similar to Hogan et al.’s findings, color was the most prominent visual variable to be mentioned. 

In terms of interface layout, the general theme for WHO was clarity. Five participants noted that WHO has a clear, spacious interface that feels lighter. For JHU, five participants found the interface crowded and cramped. One participant compared two websites with the following quote: \textit{“WHO is more clear and focused, JHU presents more data and more information to support the interpretation.”}[P3]. Another prominent visual element was the total number of cases and deaths for both websites. JHU presented the number of cases at the top left corner with red text. Three participants initially thought this was the total number of deaths instead of cases because it was written in red.
 
When explaining their actions, participants were able to describe the general interface layout when they were asked to describe what they saw and address specific interface elements during the interview in most cases. However, there were instances where some details were lost, overlooked, or confused with something else. Two participants could not remember what was on the left side, and one participant could not remember what was on the right side of JHU. During the JHU interview, One participant noted, \textit{“I click a toggle and switch to choropleth from bubble map, the color also changes to blue, I think?”}[P4]. However, the toggle they mentioned actually belongs to WHO, so they remember the blue map of WHO. Similarly, P7 said that they scrolled down to see more details of countries on JHU’s main page, which was not scrollable.

\subsubsection{Interactive Exploration}
Both WHO and JHU had an interactive map as the main element. Participants explored the data according to global events, local situations, and personal connections. Four participants supported their existing knowledge of global events from the map. An illustrative remark is: \textit{“Brazil’s cases are rising so I check it out.”}[P4]. Four participants explored the data of the country they live in. P1 noted, \textit{“We are in a 3-week trial, so I look at the UK to see if there is an increase.”}[P1]. Three participants explored the countries they have a personal connection, either the countries that they have family or friends, or planned events. An example to illustrate the latter is: \textit{“I have a planned flight for a wedding so I’m looking at both countries’ numbers.”}[P1]. Finding personal connections is a recurring theme in previous studies, Hogan et al. described this theme as “Active Seeking for Personal connections”.

We asked participants to explore the websites freely. Even though they mostly mentioned the opening pages, some of them briefly explored other pages on the websites. The page called “Critical Trends” in JHU (see Figure ~\ref{fig:Frame5}.B.3), had headlines of specific questions like “Have countries flattened the curve?” and a “Learn More” button that led to a page with a brief introduction along with a collection of related visualizations. Three participants mentioned this page, two of them assumed the “Learn more” button leads to long articles which they did not want to dive in. P3 stated that the headlines made them curious, yet they did not click the button.

During their exploration, three participants stated that they compare the information they see with what they already know. This theme of “Using previous knowledge” has been recurring in both Hogan et al.’s and Nowak et. al.’s work. Lastly, two participants stated that they might browse the website (WHO) superficially because they saw it before.

Participants expressed that they finish their exploration when they have a general idea, they think they spent enough time or when they are bored. Five participants finished their exploration when they thought they understood the general message. Two participants were finished based on time. One participant noted: \textit{“I feel like I can go on forever on this site.[...] At a point, I decided to stop myself from spending too much time.”}[P3].  Lastly, two participants stated that they finish when they start to feel bored.

\subsubsection{Information Overload}
At different stages of exploration, all participants avoided detailed analysis either to avoid confusion, out of disinterest or avoiding spending too much time. Except for P5, all participants only briefly viewed pages other than the main pages. One participant expressed: \textit{“I don’t want to distract myself, so I stayed on the main page mostly.”}[P3]. On the other hand, P5 explored the “Explorer” page on WHO in detail, yet they noted, \textit{“Too much control confused me.”} about their experience on that page. Two participants found the bar graphs on WHO, to be too detailed. P3 also noted they deliberately refrained from searching while exploring WHO because they don’t want to confuse themselves. P6 expressed their disinterest saying that they explored both websites briefly since they don’t normally look at data or statistics.

Another common theme was about participants’ negative feelings towards seeing total numbers of cases or deaths. Five participants expressed that numbers don’t mean anything to them, either because they are bad with numbers, they don’t trust the sources, or they don’t have a frame of reference. P2 expressed their need to see other information to form a frame of reference with the following quote: “ I think numbers don’t give any information, I cannot make a comparison without knowing how many people die from other diseases every day”. Two participants expressed that they prefer looking at trends rather than total numbers. To illustrate their reasoning, one participant noted: \textit{“I haven’t shown much interest in total numbers since I know that countries report them differently, so they can’t be fully trusted, I trust trends more.”}[P1].

In terms of timing, our interviews were conducted around the end of the quarantine period. Some participants expressed their negative feelings about the subject. P1 and P3 expressed that they feel tired because of the subject. P2 noted that they don’t feel curious about the subject because it affects them negatively.

\subsubsection{Expectations of Functionality}
The map was one of the first elements that participants saw and interacted with. However, five participants had negative experiences with maps on both websites. On WHO, P7 was annoyed because they could not zoom in using the touchpad, and four other participants felt disappointment and fatigue while using JHU’s map (See Figure ~\ref{fig:Frame5}.B). Three participants found zooming in and out on JHU, to be laborious. An illustrative quote is: \textit{“The dots overlap and the visual mess pushes me away, I can’t understand unless I constantly zoom in and out, and that is tiring.”}[P1]. Another participant who expected more functionality from the map noted: \textit{“Map wasn’t interesting to me, I looked to the left and right to display different data on the map but nothing happened.”}[P5]. Additionally, there were remarks about the general functionality of both websites. Three participants commented on the long load time of JHU, expressing that this caused impatience and boredom. 

\subsection{Observations and Post-Interview Questions}
\subsubsection{Observations}
Non-verbal cues are important indicators to understand if the participants are in a mindset for evocation ~\cite{hogan2015elicitation}. In a physical setting, participants breaking eye contact to look away is considered to be an indicator of them remembering their actual lived experiences, and not creating new details that are not part of the experience ~\cite{hogan2015elicitation}. Hogan et al. even suggest the researcher not to sit right across the participant to leave an empty space they can gaze at. In remote interviews, eye contact works differently than physical interviews. There is no direct eye contact since both parties look at their screens to see each other, not the camera. Previous studies on remote interviews  ~\cite{iacono2016skype} reported that this difference did not cause major problems. We also observed that it did not cause problems for our study. Participants consistently looked away from the screen, when they were remembering their experiences, so we were able to identify the evocation mindset easily.

Secondly, we asked the participants to use the websites from their laptop or desktop computers since we wanted to focus on the web experience, not mobile. All participants also joined interviewed sessions from their laptops. So, data exploration and interviews happened through the same medium, laptop computer. Two participants asked if they could open the websites during the interview, out of the frustration of not remembering details, but we asked them not to. After the interview, one other participant noted: “I felt the urge to just open the websites to remember”. We observed that using the same medium, laptop computers, both for the data exploration experience and the interview might have hindered the evocation mindset. 

Lastly, we chose a heterogeneous group of participants with varying levels of introspective practices. These ranged  from having no prior experience (1 participant), to some brief experiences (2), to regularly practicing meditation (2), to having professional background of educating others about mindfulness (1). Even though a study on micro-phenomenology advises to select participants having experience in introspective practices~\cite{nowak2018micro}, we did not observe major differences in terms of the depth or length of interviews. The interviews were not overly deep, generally, yet they created useful insights to identify reoccurring patterns and other interesting aspects of user experience.

\subsubsection{Post Interview Questions}
After finishing the elicitation interviews, we asked participants additional questions on their thoughts about the interview being remote and their other remote social experiences. We also asked what type of information sources they have followed during the COVID-19 pandemic process. 

In terms of their thoughts about the interview being remote, all participants found remote interviews to be similar or better than an interview in a physical setting. Two participants said that at the beginning of the COVID-19 process, they felt awkward or nervous, but they are very comfortable with remote experiences now. One participant stated that they even feel safer in a remote setting.

When asked about their remote social experiences during the COVID-19 process, all participants expressed that they had remote social experiences for both work and personal life. Two participants already worked remotely before the pandemic but the rest of them started to continue their work meetings, job interviews, and educations remotely after the pandemic. In terms of social life, all participants expressed that after the pandemic, they started using remote settings for socialization. Participants participated in a wide range of social activities that included small family events, webinars, movie events, yoga or musical instrument classes, and virtual nightclubs. One participant stated that they feel work and social life are intertwined.

All participants followed a source of data during the COVID-19 Pandemic. These included infographics from news, social media posts, and data visualization websites. All participants stated that they were searching for information more frequently when the pandemic started. Four of the participants previously visited WHO and two of them visited JHU.

\section{Discussion}
\subsection{Elicitation interviews give rich insights for complex interactive visualizations}
Both WHO and JHU are widely used data visualizations. Yet, our study revealed user experience issues related to both of them. The findings from the interviews include emotional responses, such as where participants felt negative about the systems. While evaluating interactive visualizations with EI is laborious (cf. \cite{nowak2018micro}), we found it to be useful for uncovering important aspects of the user experience of exploring interactive data visualizations.

Our finding that color is the primary visual element participants tried to interpret, is in line with that of previous studies \cite{hogan2015elicitation}. In addition, our study reveals color has the power to evoke an optimistic or pessimistic feeling about the subject, at first glance. Layout and visual hierarchy were two further visual aspects that were mentioned as highly relevant. Frequently, participants remarked how the spatial positioning of visual elements affected their impression of clarity or crowdedness. One important difference to previous studies was the higher level of interactivity and complexity of the examined visualizations. Even though all participants used the websites directly before the interviews, some of them had trouble remembering all details. While this can be seen as an interesting insight revealing which parts of the visualization are more and which are less memorable, we suggest future studies to combine EI with additional data collection methods like screen capturing or activity logging in order to mitigate this loss of details.

Another common theme was how participants explored the map to confirm certain aspects of global pandemic events, learn about local effects, and look into data about personally relevant places. All these relate to how people contextualize data based on their personal experiences, which corroborate similar findings of previous studies by Hogan et al. and Nowak et al. In addition, the use of color, style and level of geospatial granularity were important factors driving the users towards, or away from using the map views.

\subsection{Conducting elicitation interviews remotely}
All of our participants visited the websites from their homes, and joined the interviews remotely. This setting fits the recommendation for conducting Eliciation interviews in a quiet place, as well as using the same environment where the original experience happened. All participants stated they were comfortable with remote social settings for both professional and personal activities, especially since the pandemic. All participants expressed that it was not overly different from a physical setting for the context of the study, and some even expressed their feeling of heightened safety and comfort.

Hogan et al. suggest that showing the data visualization under investigation during the interview part might introduce bias or post rationalization. Therefore, we instructed the participants to rely on their memories during the interview. Still, some mentioned their urge to open the websites during the interview when they could not remember certain details. The utilized setting with participants using the same medium (i.e. their laptop) for experiencing and for the interview might hinder reaching the state of evocation fully. In such settings if the participant is distracted or frustrated, we suggest bringing the participant's attention back to their experience by using the questions to induce evocation state from Hogan et al.’s interview guide\cite{hogan2015elicitation}. In future studies, it might be worthwhile to investigate how the use of different mediums (such as a laptop for the experience, and a mobile phone for the interview) affects the outcome.

At the recruitment stage, we initially tried to set up a date for interviews with all participants. From these ten participants, five participants rescheduled once and six participants requested to be interviewed spontaneously. In the end, three participants canceled the interview due to their schedule and we ended up interviewing the remaining seven. While we can not compare cancellation and rescheduling rates to physical interviews, we believe that remote interviews enabled more room for leeway. This might have negative aspects when the study employs additional more structured methods, or when multiple interviewers need to coordinate. However, if the study design is flexible and open to spontaneity, remote interviews can allow adjusting to participants’ schedules, therefore requiring less effort for them, and thus resulting in a more accommodating environment.

Lastly, for the case of investigating interactive visualizations for the general public, we observed no major differences between participants with or without prior introspective practices. In any case, it is important to communicate that the aim of the interview is not about the participant’s performance but about bringing attention to the sensory, emotional, and hedonistic aspects of the experience which might stay unnoticed during the experience itself.

\subsection{Visualizing complex pandemic data for a global audience}

Investigating the subjective experience people have with visualizations has been identified as one major evaluation scenario~\cite{lam2011empirical}. These types of studies are critical in order to understand the personal relations people have with data visualizations that affect us all. 

We observed that the choice of colors and design can greatly impact users’ perception of the topic even when the data is the same or very similar. In general, participants found the WHO to paint a positive picture while JHU created a sense of danger. We conducted the interviews in June 2020, and most of our participants have been in self-quarantine for several months. The pessimistic feeling JHU evoked disturbance for the participants who wanted to see some positivity. One participant suggested that the look and feel of such a visualization should be different at the beginning of the pandemic, where it is more important to draw attention to the severity of the subject and through the end it should have a more positive look and feel to create hope. 

We observed that participants were prone to be easily overwhelmed by the data. During the interviews, they limited their time and mostly refrained from deeper analysis. When they engaged in deeper analysis, too much control in adjusting the visualized data created confusion (e.g. “Explorer” page in WHO, see Figure~\ref{fig:Frame5}.A.2). We learned that it is useful to make the main takeaway very clear. A good example is the summarizing sentence at the bottom of WHO (see Figure~\ref{fig:Frame5}.A.1), which helped users to grasp the general situation easily. Lastly, many participants stated that they do not fully trust numbers. Instead of making numbers a strong visual element as it is common in dashboards, trend lines and descriptive text such as well designed labels and help texts that support interpreting data can be helpful for users.

\section{Conclusion}
In this paper, we presented a case study on remotely gathering subjective feedback of people's experience with interactive data visualizations in the context of the coronavirus pandemic. Elicitation Interviews aim to reveal hedonistic, sensory, and emotional aspects of using data visualizations, with previous studies mostly focusing on static or narrative visualizations~\cite{hogan2015elicitation,nowak2018micro}. We conducted remote elicitation interviews with two major data-driven dashboards. The two systems in our study were complex interactive visualizations about the COVID-19 pandemic data, a global and current topic that affects many people on a personal level. Therefore we believe it is important to understand the user experience beyond performance measures and explore the personal driving factors of open-ended exploration. 

Our findings indicate that spatial exploration can be driven by a user's existing knowledge about global events, curiosity about the user's local situation, and personal connections with places. For both visualizations, users were commonly interested in the overall picture and refrained from deeper analysis. Users also generally preferred seeing trends over absolute numbers of cases and deaths. Supporting previous findings, we identified color as an important element that enables users to interpret data. Furthermore, we observed that color can induce positive or negative feelings towards the visualized topic.

In terms of conducting Elicitation Interviews remotely, our observations indicate that users felt safe and comfortable with remote interviews due to their increased experience in remote activities for work, school, and social life during the pandemic. However, we also noticed that sustaining an evocation state can be challenging in remote settings. Our work is a step towards understanding the human experience with data more thoroughly, in settings where conducting in-person studies are not possible. We aim to investigate combining remote elicitation interviews with additional objective methods like activity logging or screen capturing in the future.

\acknowledgments{
We wish to thank all participants that volunteered for our study.}

\bibliographystyle{abbrv-doi}

\bibliography{template}
\end{document}